\documentclass[draft,numberedheadings]{aipproc}

\layoutstyle{8x11single}

\begin{document}

\def\simg{\mathrel{\rlap{\raise 0.511ex \hbox{$>$}}{\lower 0.511ex \hbox{$\sim$}}}}
\def\siml{\mathrel{ \rlap{\raise 0.511ex \hbox{$<$}}{\lower 0.511ex \hbox{$\sim$}}}}
\parskip 4pt

\title{Gamma-Ray Burst afterglows: theory and observations}


\author{A. Panaitescu}{ address={ ISR-1, Los Alamos National Laboratory, Los Alamos, 
   NM 87545, USA} }

\begin{abstract}
\\
 I discuss some theoretical expectations for the synchrotron emission from a relativistic 
blast-wave interacting with the ambient medium, as a model for GRB afterglows, and compare 
them with observations. An afterglow flux evolving as a power-law in time, a bright optical 
flash during and after the burst, and light-curve breaks owing to a tight ejecta collimation 
are the major predictions that were confirmed observationally, but it should be recognized
that light-curve decay indices are not correlated with the spectral slopes (as would be 
expected), optical flashes are quite rare, and jet-breaks harder to find in Swift X-ray 
afterglows. 

 The slowing of the early optical flux decay rate is accompanied by a spectral evolution,
indicating that the emission from ejecta (energized by the reverse shock) is dominant in 
the optical over that from the forward shock (which energizes the ambient medium) only up to 
1 ks. However, a long-lived reverse shock is required to account for the slow radio flux 
decays observed in many afterglows after $\sim 10$ day.

 X-ray light-curve plateaus could be due to variations in the average energy-per-solid-angle 
of the blast-wave, confirming to two other anticipated features of GRB outflows: energy 
injection and angular structure. The latter is also the more likely origin of the fast-rises 
seen in some optical light-curves. To account for the existence of both chromatic and achromatic 
afterglow light-curve breaks, the overall picture must be even more complex and include
a new mechanism that dominates occasionally the emission from the blast-wave: either late 
internal shocks or scattering (bulk and/or inverse-Compton) of the blast-wave emission
by an outflow interior to it. 

\end{abstract}

\maketitle

\section{Introduction}

 A relativistic motion of GRB sources was advocated by \cite{paczynski86,goodman86} from
that the energies released exceed by many orders of magnitude the Eddington luminosity for
a stellar-mass object, especially if GRBs are at cosmological distances (see also 
\cite{shemi90,meszaros92}).

 The detection by CGRO/EGRET of photons with energy above 1 MeV during the prompt burst 
emission (e.g. \cite{hurley94}) shows that GRB sources are optically thin to such photons.
Together with the sub-MeV burst isotropic-equivalent output of $10^{52}-10^{54}$ ergs 
(e.g. \cite{bloom01}) and the millisecond burst variability timescale, the condition for
optical thickness to high energy photons gives another reason why GRBs must arise from 
ultra-relativistic sources, moving at Lorentz factor $\Gamma \simg 100$ 
(e.g. \cite{fenimore93,lithwick01}).

 The same conclusion is enforced by the measurement of a relativistic expansion of the
radio afterglow source. That expansion was either measured directly, as for GRB 030329 
($z=0.17$), whose size increased at an apparent speed of 5c, indicating a source expanding 
at $\Gamma \siml 6$ at 1--2 months \cite{taylor04}, or was inferred from the rate at which 
interstellar scintillation \cite{goodman97} quenches owing to the increasing source size, 
as for GRB 970508, whose expansion speed is inferred to be close to $c$ at 1 month
\cite{waxman98}. The adiabatic dynamical evolution of a blast-wave, $\Gamma^2 M = const$, 
where $M$ is the mass of the ambient medium, leads to $\Gamma \propto t^{-3/8}$ for a 
homogeneous medium ($t$ being the observer-frame photon arrival time). Then, $\Gamma 
(30\,{\rm d}) = 2$ extrapolated to the burst time implies $\Gamma(100\,{\rm s}) \sim 100$. 
Extrapolating to such early times is justified by that most optical afterglow light-curves 
display a power-law decay starting after the burst, which sets a lower limit on the source 
Lorentz factor at that time.

\begin{figure}
    \parbox[h]{7cm}{ \includegraphics[height=6.4cm,width=7cm]{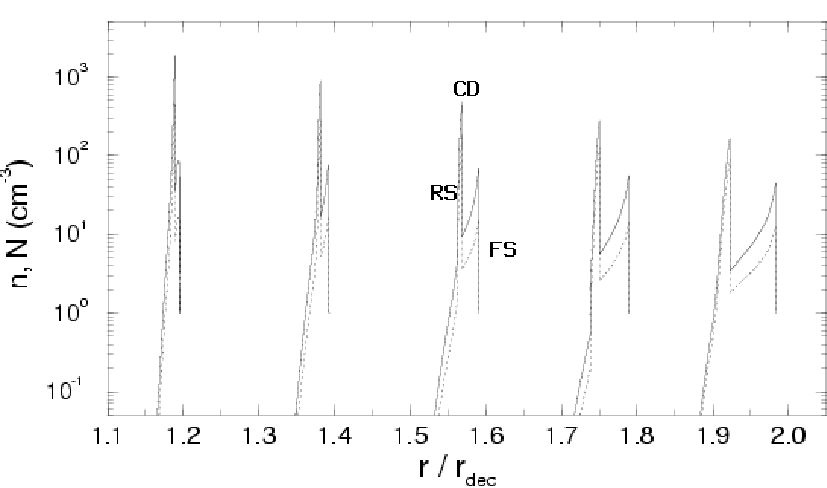} }
    \hspace*{2mm}  
    \parbox[h]{8cm}{ \includegraphics[height=6cm,width=8cm]{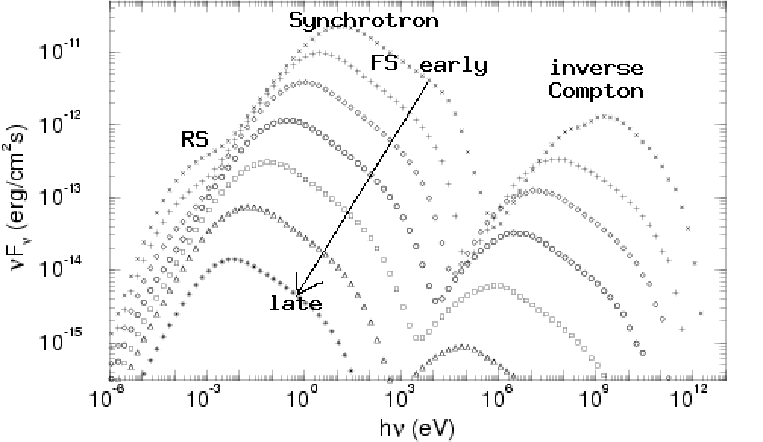} }
\caption{ \footnotesize
  {\sl Left panel}: particle density distribution in a relativistic blast-wave (moving 
  from left to right) sweeping the ambient medium (of particle density $1\,{\rm cm^{-3}}$)
  located ahead of the forward shock (FS). The ejecta are located behind the contact 
  discontinuity (CD) and are energized by the reverse shock (RS). Solid lines are for
  lab-frame density, dotted lines show comoving density. Locations are measured in 
  deceleration radii, defined by the first reverse shock crossing the ejecta shell
  (other reverse shocks can develop later). 
  {\sl Right panel}: as the blast wave decelerates, the synchrotron and inverse-Compton 
  emissions from the reverse and forward shocks become weaker and softer (assuming
  constant electron and magnetic fields parameters). For a brief release of the ejecta,
  the reverse shock is only mildly relativistic, radiates at lower frequencies, and 
  accelerates electrons for a shorter duration (while it crosses the ejecta shell), 
  after which electrons cool adiabatically and their characteristic synchrotron frequency 
  may fall below that of observations. 
  Results shown here are based on 1-dimensional hydrodynamical simulations of the 
  ejecta--medium interaction \cite{wen97}. }
\label{bw}
\end{figure}

 Whether the GRB ejecta are a cold baryonic outflow accelerated by the adiabatic losses of 
fireball's initial thermal/radiation energy (e.g. \cite{meszaros93,piran93}), or relativistic 
pairs formed through magnetic dissipation in a Poynting outflow, as in the electromagnetic 
model of \cite{lyutikov06}, their interaction with ambient medium will drive two shocks: 
a reverse shock crossing the ejecta and a forward-shock sweeping the circumburst 
medium, as illustrated in Figure \ref{bw}. Both shocks energize their respective media, 
accelerate relativistic particles and generate magnetic fields through some plasma-instability
related process, such as the two-stream Weibel instability driven by an anisotropic particle
distribution function \cite{medvedev99}. The original magnetic field of the fireball at 
$10^7$ cm becomes too weak by the time the fireball reaches the $10^{15}-10^{17}$ cm radius 
(where the burst and afterglow emissions are produced) for the synchrotron emission to account 
for the sub-MeV burst emission and for the longer-wavelength, ensuing afterglow emission,
even if the fireball was initially magnetically dominated \cite{meszaros93,medvedev99}.

\begin{figure}
    \parbox[h]{8cm}{ \includegraphics[height=6cm,width=8cm]{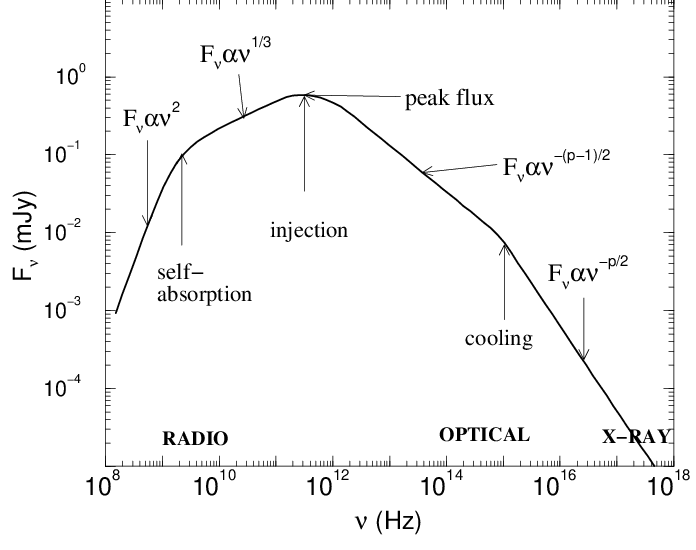} }
    \hspace*{2mm}
    \parbox[h]{8cm}{ \vspace*{-5mm} \footnotesize
     Broadband spectrum of the synchrotron emission showing its three break frequencies: 
     self-absorption (in radio), injection ($\nu_i$, corresponding to the energy of the 
     electrons currently accelerated by the shock), and cooling ($\nu_c$, for electrons whose 
     radiative cooling timescale is equal to the dynamical timescale). 
     The energy distribution of the shock-accelerated electrons is assumed to be a power-law
     above the injection break -- $dN_e/d\epsilon \propto \epsilon^{-p}$ -- which leads
     to a power-law afterglow spectrum ($F_\nu \propto \nu^{-\beta}$) of slope $\beta = 
     (p-1)/2$ between $\nu_i$ and $\nu_c$ and $\beta = p/2$ above $\nu_c$.
     For a forward-shock afterglow origin, the three break frequencies and the peak flux 
     determined from a "snap-shot" afterglow spectrum provide four constraints for four 
     unknown parameters: ejecta kinetic energy (per solid angle), ambient density, and 
     fractional energies of electrons and magnetic field. Thus, afterglow spectra extending 
     down to radio frequencies can be used to determine afterglow fundamental physical 
     parameters (e.g. \cite{wijers99,granot99}). }
\caption{}
\label{spek}
\end{figure}

 The evolution of the synchrotron and inverse-Compton fluxes produced by the blast-wave at
a fixed frequency (i.e. the light-curve) is determined by how the characteristics of the
spectrum (break frequencies and peak flux) change with time. Figure \ref{spek} shows the
expected afterglow synchrotron spectrum, whose characteristics depend on the blast-wave radius, 
number of radiating electrons, their distribution with energy, magnetic field strength, 
and Lorentz factor. 

 If the typical electron energy and magnetic field energy correspond to some fixed fraction 
of the post-shock energy, or if they start from such a fixed fraction and then evolve 
adiabatically (as for adiabatically colling ejecta), then the afterglow light-curve depends 
on 
(1) the evolution of the blast-wave Lorentz factor, the blast-wave radius being 
    $R \simeq \Gamma^2 ct$ (with $t$ the photon-arrival time measured since burst trigger)
(2) the spectrum of the blast-wave emission (i.e. the distribution of electrons with energy), 
    and, in the case of the reverse-shock, 
(3) the evolution of the incoming mass.

 The power-law deceleration of the blast-wave ($\Gamma \propto r^{-(3-s)/2} \propto 
t^{-(3-s)/(8-4s)}$ for $s<3$, where $n \propto r^{-s}$ is the radial stratification of the 
ambient medium density) and the power-law afterglow spectrum ($F_\nu \propto \nu^{-\beta}$) 
are two factors which lead to a power-law afterglow light-curve ($F_\nu \propto t^{-\alpha}$), 
with the decay index $\alpha$ being a linear function of the spectral slope $\beta$. 
 These are the only two factors at work for the forward-shock emission and the ejecta emission 
during the adiabatic cooling phase (which starts when the reverse shock has crossed the ejecta 
shell), the two models that yield power-law afterglow light-curves in the most simple and
natural way.

 In contrast, for the reverse-shock emission (i.e. the ejecta emission while the shock exists), 
the light-curve depends also on the radial distribution of ejecta mass and of their Lorentz 
factor, thus the observed
power-law light-curves require additional properties to be satisfied by the relativistic
ejecta. Such properties seem {\sl ad-hoc} when it comes to explaining single power-law 
afterglows whose flux displays an unchanged decay over 2--4 decades in time (such as the 
X-ray afterglows of GRBs 050801, 050820A, 06011B, 060210, 060418, 061007), but they also 
provide the flexibility required to account for the prevalent X-ray afterglow light-curves 
that exhibit one or more breaks.

\section{Afterglow light-curves}

 I consider first the afterglow emission at early times, when the blast-wave is sufficiently 
relativistic that the observer receives boosted emission from a region of half-angle opening 
$\Gamma^{-1}$ (as seen from the center of the blast-wave) that is smaller than the half-aperture 
$\theta_{jet}$ of the collimated outflow. In that case, the observer does not "see" yet the
angular boundary of the outflow and the received emission is as bright as for a spherical
blast-wave. The evolution of the spectral characteristics of the emission from ejecta and the
swept-up ambient medium are presented below, the resulting power-law decay indices of the
synchrotron flux being listed in Table 1.

\subsection{Ejecta emission (energized by reverse shock or cooling adiabatically)}

 For a short-duration ejecta release, the reverse shock crosses the ejecta shell over 
an observer-frame time that depends primarily on the ejecta Lorentz factor $\Gamma_0$: 
$t_{dec} = 350\, (z+1) (E_{53}/n_0\Gamma_{0,2}^8)^{1/3}$ s, where $E_{53}$ is the 
isotropic-equivalent ejecta kinetic energy in $10^{53}$ erg, $n_0$ is the ambient medium 
density in protons per ${\rm cm^3}$, and $\Gamma_{0,2} = \Gamma_0/100$. In this case, 
the reverse shock is semi-relativistic and, more likely, radiates below the optical. 
After $t_{dec}$, the input of energy into the shocked structure ceases and the blast-wave 
begins to decelerate.

 If the ejecta release is an extended process, the deceleration timescale depends primarily
on the duration $\tau$ over which the ejecta are expelled: $t_{dec} = 0.7\,(z+1) \tau$.
In this case, the reverse shock is relativistic and could produce a bright optical emission.
The separation between these two cases is set by $t_{dec}(\Gamma) = t_{dec}(\tau)$, 
"short-duration ejecta release" meaning $\tau < \tilde{\tau} \equiv 500\, (E_{53}/n_0
\Gamma_{0,2}^8)^{1/3}$ s. 

 For a wind medium, the deceleration timescale is $t_{dec} = 3\, (z+1) E_{53}/ (A_* 
\Gamma_{0,2}^4)$ s for a short-duration ejecta release, where $A_*$ is the wind density
parameter, normalized to that resulting for $10^{-5}\, M_\odot/{\rm yr}$ being ejected
at a terminal velocity of 1000 km/s, $t_{dec}$ being the same as for a homogeneous medium
in the case of a long-duration ejection (which occurs for $\tau > \tilde{\tau} \equiv 4 
\, E_{53}/(A_* \Gamma_{0,2}^4)$ s).

 At the deceleration radius, $\sim 1/3$ of the ejecta energy has been transferred
to the swept-up ambient medium, which moves at $\Gamma \simeq (2/3) \Gamma_0$ for 
$\tau < \tilde{\tau}$ and a lower $\Gamma$ for $\tau > \tilde{\tau}$. Taking into account 
that the energy per particle in the post forward-shock gas is $\Gamma$, it follows that,
at $t_{dec}$, the ejecta mass is larger than that of the forward shock by at most
a factor $\simeq \Gamma_0$. This implies that, at $t_{dec}$, the peak flux of the 
reverse-shock emission spectrum is a factor $\simg 100$ larger than the peak flux
of the forward-shock spectrum, hence, the optical flash from the reverse shock could
be up to 5 magnitudes brighter than the optical emission from the forward shock.

 As mentioned above, when there is a reverse shock crossing the ejecta, its emission
flux should depend on the density and Lorentz factor of the incoming ejecta. 
Semi-analytical calculations of the ejecta synchrotron emission when there is a reverse 
shock have been done by \cite{uhm07} and \cite{genet07} for a density and Lorentz factor 
of the incoming ejecta tailored to produce X-ray light-curve plateaus, hydrodynamical 
calculations of the reverse-shock dynamics have been presented by \cite{kobayashi00}, and 
calculations of the ejecta emission after the reverse shock has crossed the ejecta (i.e. 
during adiabatic cooling) have been published by \cite{meszaros97,sari99,kobayashi00A,alin04}. 
However, analytical calculations of the ejecta emission while there is a reverse shock 
\cite{kobayashi03} have yet to be done.

 Assuming a uniform ejecta density \& Lorentz factor and an extended ejecta release,
for which the reverse shock is relativistic and the shocked ejecta are
slightly decelerated even before $t_{dec}$, owing to the progressive dilution of the
incoming ejecta, I find that the {\sl reverse-shock} peak flux $F_p$ and break frequencies
$\nu_i$ (injection) and $\nu_c$ (cooling) evolve as $F_p \propto t^{-1/5}$, $\nu_i \propto
t^0$, $\nu_c \propto t^{-4/5}$ for a homogeneous medium and $F_p \propto t^{-1/3}$, 
$\nu_i \propto t^{-2/3}$, $\nu_c \propto t^{2/3}$ for a wind.

 As for the {\sl ejecta} emission decay during the adiabatic cooling phase, the evolution 
of the spectral characteristics is approximately $F_p \propto t^{-0.67}$, 
$\nu_i,\nu_c \propto t^{-1.2}$ for a homogeneous medium, and $F_p \propto t^{-0.80}$, 
$\nu_i,\nu_c \propto t^{-1.5}$ for a wind. For $\nu_{obs} < \nu_i$, the cooling ejecta
flux should decay with an index $\alpha = 0.3$, for either type of medium. Above $\nu_i$
but below $\nu_c$, the decay index is $\alpha = 1.19 \beta - 0.67$ for a homogeneous medium
and $\alpha = 1.47 \beta - 0.80$ for a wind. Depending on the treatment of the ejecta 
dynamics and adiabatic cooling, other researchers reached slightly different results 
$\alpha = a\beta + k$ with
$a \simeq 1.3$ and $k \simeq 0.8$ in \cite{meszaros97},
$a \simeq 1.5$ and $k \simeq 1.0$ in \cite{sari99}, 
$a \in (1.37,1.66)$ and $k \in (0.75,0.96)$ in \cite{kobayashi00A}.

 After $\nu_c$ falls below $\nu_{obs}$ owing to adiabatic cooling, the observer receives 
no emission from the area of angular opening $\Gamma^{-1}$ moving directly toward the 
observer (because of the exponential cut-off of the synchrotron emissivity above the 
synchrotron peak) but receives emission from the fluid moving at increasing angles larger 
than $\Gamma^{-1}$. That emission (called {\sl large-angle} emission, lacking a better name) 
was released at the same time as the emission from angles less than $\Gamma^{-1}$, but 
arrives later at observer because of the spherical curvature of the emitting surface 
and finite speed of light, and is less beamed relativistically. As shown by \cite{kumar00}, 
the large-angle emission is characterized by $F_p \propto t^{-2}$, $\nu_i,\nu_c \propto
t^{-1}$, the power-law decay index being $\alpha = \beta + 2$ (see also \cite{fenimore97})
These are general results, arising only from relativistic effects, and independent 
of the emission process. The only assumption made in its derivation is that the surface 
emissivity properties are angle-independent.

\subsection{Forward-shock emission (energized ambient medium)}

 Before deceleration of the blast-wave begins, the shocked ambient medium moves at a constant 
$\Gamma \simeq (2/3)\Gamma_0$, if the reverse shock is semi-relativistic, or is slowly 
decelerating as $\Gamma \propto t^{-(3-s)/(10-2s)}$, if the reverse shock is relativistic. 

 For a {\sl semi-relativistic} reverse shock, the spectral characteristics of the
{\sl pre-deceleration} forward-shock synchrotron emission evolve as 
$F_p \propto t^3$, $\nu_i = const$, $\nu_c \propto t^{-2}$ for a homogeneous medium, and 
$F_p = const$, $\nu_i \propto t^{-1}$, $\nu_c \propto t$ for a wind. 
 For a {\sl relativistic} reverse shock, the above scalings become
$F_p \propto t^{3/5}$, $\nu_i \propto t^{-6/5}$, $\nu_c \propto t^{-4/5}$ for $s=0$, and 
$F_p \propto t^{-1/3}$, $\nu_i \propto t^{-4/3}$, $\nu_c \propto t^{2/3}$ for $s=2$.

 The forward-shock synchrotron emission {\sl after deceleration} has received the most
attention (e.g. \cite{meszaros97,sari98,alin98,chevalier99,granot99,granot02}). Under the 
usual assumptions of constant blast-wave energy and micro-physical electron and magnetic 
field parameters, the forward-shock peak flux and spectral break frequencies evolution is
$F_p \propto t^0$, $\nu_i \propto t^{-3/2}$, $\nu_c \propto t^{-1/2}$ for a homogeneous medium, 
and $F_p \propto t^{-1/2}$, $\nu_i \propto t^{-3/2}$, $\nu_c \propto t^{1/2}$ for a wind.
From here, it follows that the flux below $\nu_i$ should rise slowly as $F_\nu \propto 
t^{1/2}$ for a homogeneous medium or be constant for a wind medium.
For $\nu_i < \nu_{obs}$, the forward-shock flux decay is a power-law of index $\alpha = 
(3/2)\beta + k$, with
$k=1/2$ if ambient medium has a wind-like stratification (as expected for a massive 
stellar long-GRB progenitor) and if $\nu_{obs} < \nu_c$,  
$k=0$ for a homogeneous medium (which, surprisingly, is more often found to be compatible 
with the observed afterglows than a wind) if $\nu_{obs} < \nu_c$, and 
$k=-1/2$ if $\nu_c < \nu_{obs}$, for any type of medium.


 All the above results hold for a spherical outflow or a collimated one before the jet 
boundary becomes visible to the observer. At the jet-break time $t_{jet}$, when deceleration 
lowers the jet Lorentz factor to $\Gamma = \theta_{jet}^{-1}$, the emission from the jet edge 
is no longer relativistically beamed away from the direction toward the observer. 
At $t > t_{jet}$, the lack of emitting fluid at angles larger than $\theta_{jet}$ leads to 
a steepening of the afterglow decay by $\Delta \alpha = 3/4$ for a homogeneous medium and 
$\Delta \alpha = 1/2$ for a wind. 
Simultaneously, the lateral spreading of the jet becomes important and 
leads to a faster deceleration of the jet, which switches from a power-law in the blast-wave 
radius to an exponential \cite{rhoads99}, yielding an extra steepening of the afterglow decay
of magnitude smaller or comparable to $\Delta \alpha$ above. Together, these two {\sl jet 
effects} lead to $F_p \propto t^{-1}$, $\nu_i \propto t^{-2}$, $\nu_c \propto t^0$, rather
independent of the ambient medium stratification, and a post jet-break forward-shock flux 
decay of index $\alpha = 1/3$ below $\nu_i$, while for $\nu_{obs} > \nu_i$, one obtains 
$\alpha = 2\beta + 1$ below $\nu_c$ and $\alpha = 2\beta$ above $\nu_c$.

\begin{table}
\begin{tiny} 
\caption{\footnotesize
    Index $\alpha$ of the power-law decay of a spherical blast-wave synchrotron flux 
    ($F_\nu \propto t^{-\alpha}$) for various models:
    RS(1) = {\sl reverse-shock}, assuming an ejecta shell of uniform density and Lorentz factor;
    RS(2) = adiabatically cooling {\sl ejecta};
    FS(1) = pre-deceleration {\sl forward shock} for a semi-relativistic reverse shock,
    FS(2) = pre-deceleration {\sl forward shock} for a relativistic reverse shock;
    FS(3) = decelerating {\sl forward shock}. 
    $\beta$ is the slope of the spectrum ($F_\nu \propto t^{-\beta}$) at the observing frequency 
    $\nu$.  }
\begin{tabular}{|l|lllll|lllll|}
 \multicolumn{11}{l}{} \\
 \hline 
   & \multicolumn{4}{c}{HOMOGENEOUS MEDIUM} & &  \multicolumn{4}{c}{WIND MEDIUM} & \\ 
 \hline
  MODEL
   & $\nu<\nu_i<\nu_c$&$\nu<\nu_c<\nu_i$&$\nu_i<\nu<\nu_c$&$\nu_c<\nu<\nu_i$&$\nu_{i,c}<\nu_i$ 
   & $\nu<\nu_i<\nu_c$&$\nu<\nu_c<\nu_i$&$\nu_i<\nu<\nu_c$&$\nu_c<\nu<\nu_i$&$\nu_{i,c}<\nu_i$ \\
 \hline
  RS(1) & 1/5 & 1/15 & 1/5 & 3/5 & 3/5 & 1/9 & 5/9 & $2/3\beta+1/3$ & 0 & $2\beta/3-1/3$ \\ 
 \hline 
  RS(2) & 0.3 & 0.3 & $1.2\beta+0.7$ &--& $\beta+2$ & 0.3 &  0.3 & $1.5\beta+0.8$ &--& $\beta+2$ \\ 
 \hline
  FS(1) & -3 & -11/3 & -3 &  -2 & -2 & -1/3 & 1/3 & $\beta$ & -1/2 & $\beta-1$ \\ 
 \hline
  FS(2) & -1 & -19/15 & $1.2\beta-0.6$ & -1/5 & $1.2\beta-0.4$ & -1/9 & 5/9 & $4/3\beta+1/3$ & 0 & 
    $4/3\beta-2/3$ \\ 
 \hline
  FS(3) & -1/2 & -1/6 & $1.5\beta$ & 1/4 & $1.5\beta-0.5$ & 0 & 2/3 & $1.5\beta+0.5$ & 1/4 & 
   $ 1.5\beta - 0.5$ \\ 
 \hline
\end{tabular}
\end{tiny}
\end{table}

\section{Afterglow observations}

 \cite{paczynski93} were the first to predict the existence of 
radio afterglows following the burst phase. \cite{meszaros97} have analyzed two models for 
the reverse shock emission and one for the forward shock, predicting long-lived optical 
afterglows with a flux decaying as a power of time. The first detection of an afterglow 
and measurement of a power-law flux decay followed soon (GRB 970228 \cite{wijers97}), with 
many other optical \cite{kann07} and X-ray afterglows \cite{obrien06,willingale07} having 
been observed until today.

 In general, the broadband (radio, optical, X-ray) emission of GRB afterglows display the
expected power-law spectra and light-curves, as well as other features, which, in chronological 
order of their {\sl prediction} are: radio scintillation (\cite{goodman97} \& \cite{waxman98}),
optical counterpart flashes (\cite{sari99} \& \cite{akerlof99}), jet-breaks (\cite{rhoads99}
\& \cite{kulkarni99}), dimmer afterglows for short bursts (\cite{alin01} \& \cite{kann08}). 
GRB afterglows display sufficient diversity (e.g. wide luminosity distributions at all observing 
frequencies, non-universal shock micro-physical parameters) and puzzling features (slowly-decaying 
radio fluxes, X-ray light-curve plateaus, chromatic X-ray light-curve breaks) to challenge the 
standard external-shock model and warrant various modifications. Below, I discuss some of
these issues.

\subsection{Light-curves and spectra}

 According to the temporal scalings identified in the previous section, the light-curves
of GRB afterglows should display rises in the early phase, if the forward-shock emission
is dominant (because the reverse-shock flux is most often expected to decay), followed
by a decay, both being power-laws in time. Furthermore, the broadband afterglow spectra
are expected to be rising at (radio) frequencies below the spectrum peak and fall-off
at higher (X-ray) photon energies. Also expected is that the light-curve decay indices
$\alpha$ and spectral slopes $\beta$ satisfy one or more closure relationships and that,
there is a positive correlation between $\alpha$ and $\beta$ (from that $d\alpha/d\beta 
\in [1,2]$). 

\begin{figure}
    \parbox[h]{7cm}{ \includegraphics[height=6cm,width=7cm]{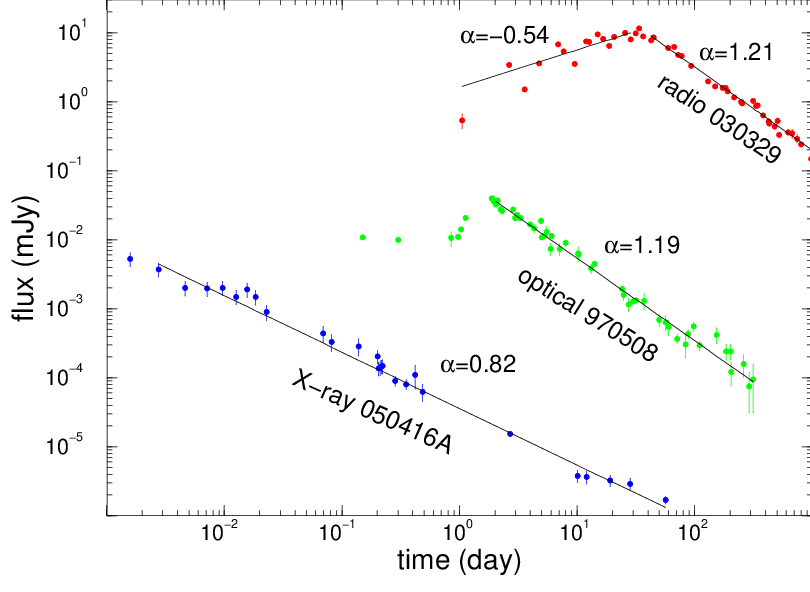} }
    \hspace*{2mm}
    \parbox[h]{7cm}{ \includegraphics[height=6cm,width=7cm]{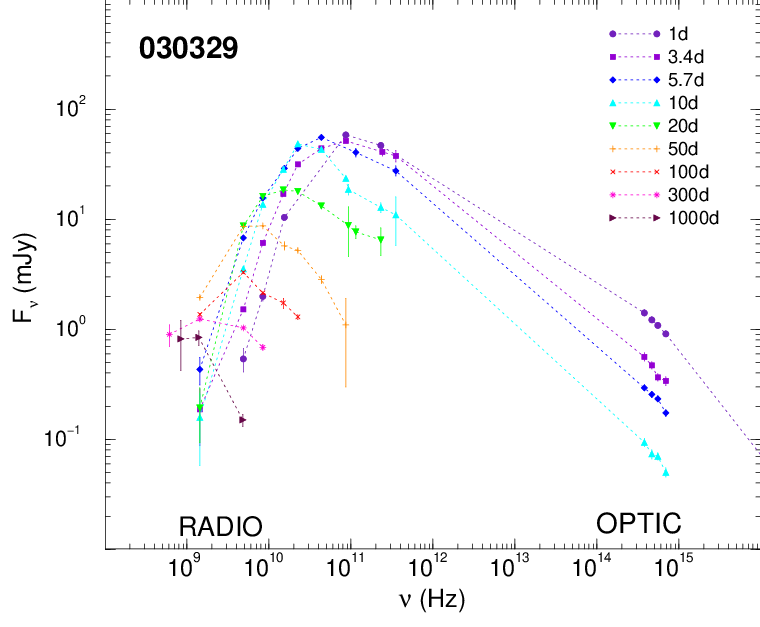} }
\caption{ \footnotesize
    {\sl Left panel}: examples of radio, optical, and X-ray afterglows displaying long-lived 
    flux power-law decays. The radio flux of GRB afterglow 030329 is affected by scintillation,
    which quenches after few tens of days. The optical light-curve of GRB afterglow 970508
    displayed an unusual, late rise at 1--2 days.
    {\sl Right panel}: evolution of the radio-to-optical spectrum of GRB afterglows 030329,
    showing its gradual softening (decreasing peak energy) and dimming (decrease of peak flux).
    Color coding used indicates earlier times with bluer colors and later times with redder 
    colors. }
\label{aglows}
\end{figure}

 Figure \ref{aglows} illustrates the flux power-law decays and power-law spectra typically
observed for GRB afterglows. The light-curves chosen there display long-lived power-law
decays, but many afterglows exhibit more diversity, their light-curves showing two or three
power-law decays, sometimes even rising at earlier times, rarely exhibiting brightening
episodes (in optical or X-ray) or sudden drops (in X-ray). 

 The broadband spectrum of GRB afterglow 030329 shown in Figure \ref{aglows} displays
an optically thick part (to self-absorption) in the radio, at earlier times, a constant 
peak flux up to 10 days, during which the radio flux rises slowly, followed by a decreasing 
peak flux and decreasing radio flux. For $\nu_{obs} < \nu_i$, as required by the radio 
spectrum (right panel), the rise of the radio flux, $F_{GHz} \propto t^{0.54}$ (left panel), 
is consistent with that expected from a decelerating forward-shock interacting with a 
homogeneous medium ($F_\nu \propto t^{1/2}$), and marginally consistent with the forward-shock 
pre-deceleration emission for either a semi-relativistic reverse shock and wind medium
($F_\nu \propto t^{1/3}$) or a relativistic reverse shock and homogeneous medium ($F_\nu 
\propto t^1$). 

 The evolution of spectral breaks is generally hard to determine observationally and use
for identifying the correct afterglow model: self-absorption affects only the early radio
emission, when the large flux fluctuations are caused by interstellar scintillation, while
the cooling break is too shallow and evolves too slowly to be well measured even if it fell
in the optical or X-ray bands. The best prospects for this test is offered by the injection
break, which should cross the radio domain at tens of days, when the scintillation amplitude
is reduced by the larger source size. 

\begin{figure}
    \parbox[h]{11.5cm}{ \includegraphics[height=5cm,width=11.5cm]{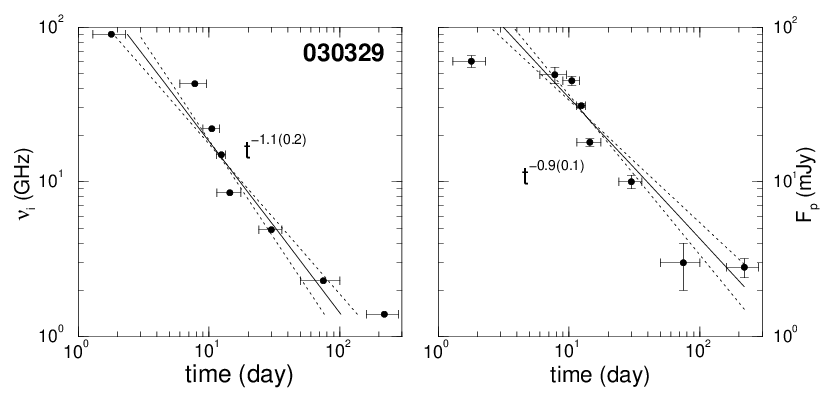} }
    \hspace*{2mm}
    \parbox[h]{4.5cm}{ \footnotesize
     Evolution of the peak frequency $\nu_i$ and peak flux $F_p$ of the radio spectrum of 
     GRB afterglow 030329, with peak time and $F_p$ measured from the peaks of radio
     light-curves (similar results would be obtained from the peaks of radio spectra at 
     various epochs -- right panel Figure \ref{aglows}). 
     Solid lines show best-fit power-laws, dotted lines indicate $1\sigma$ fits. }
\caption{}
\label{radiopk}
\end{figure}

 Sufficient radio coverage to construct radio afterglow spectra at many epochs and determine
the peak frequency $\nu_i$ and flux $F_p$ is rarely achieved. GRB 030329 is one such case
(Figure \ref{radiopk}), the evolution of $\nu_i$ being slower than expected for a spherical
blast-wave or a jet that does not expand (yet or ever) laterally (for either, $\nu_i \propto 
t^{-3/2}$ for any medium stratification), while that of $F_p$ is close to that expected for a 
jet spreading laterally (for which $F_p \propto t^{-1}$). Thus, the evolutions of $\nu_i$ and
$F_p$ for GRB afterglow 030329 seem mutually inconsistent. The slower-than-expected evolution 
of $\nu_i$ requires that shock micro-physical are not constant (as assumed in the standard
model), but the evolution of $F_p$ can be accounted for by a spherical blast-wave provided 
that the ambient medium density decreases as $n \propto r^{-2.5}$.

\begin{figure}
    \parbox[h]{7cm}{ \includegraphics[height=5.5cm,width=7cm]{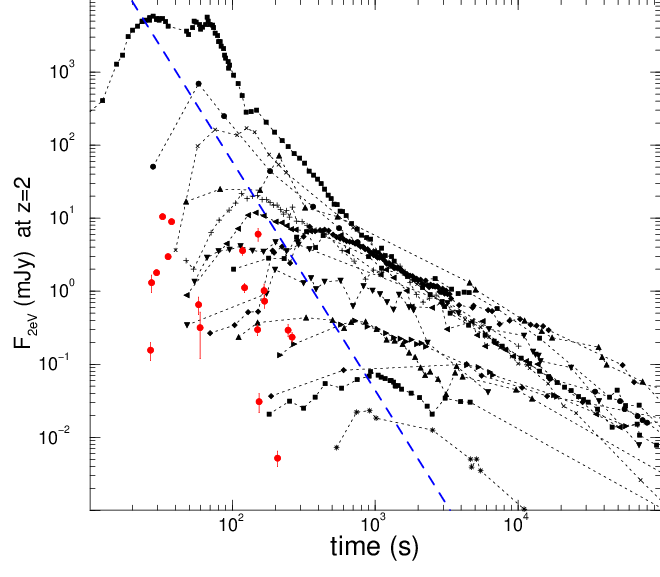} }
\hspace*{7mm}
    \parbox[h]{6cm}{ \footnotesize
    Optical light-curves of 15 afterglows displaying a rise, moved at same redshift.  
    Red symbols show the first measurement of 19 other optical afterglows that exhibit
    a decay from the beginning.
    The blue dashed line indicates the best-fit to the peak time--peak flux anti-correlation 
    of rising afterglows ($F_p \propto t_p^{-3.3 \pm 0.2}$).
    The earlier rises are sharp ($F_{opt} \propto t^{2.5 \pm 0.5}$) and would require a 
    homogeneous medium if they were attributed to the pre-deceleration emission from an 
    isotropic forward-shock. } 
\caption{}
\label{rises}
\end{figure}

 Rising optical light-curves have been seen for more than a dozen afterglows up to 1 ks after
trigger (Figure \ref{rises}). If interpreted as the due to the pre-deceleration emission 
from the forward shock (e.g. \cite{molinari07}), they require smaller than average initial 
ejecta Lorentz factors (if the reverse shock is semi-relativistic) or longer-lasting ejections. 
The existence of energetic ejecta with a significantly smaller ejecta Lorentz factor could 
also explain the late (1--2 d) sharp rise displayed by the optical afterglow of GRB 970508 
(Figure \ref{aglows}).

 However, late-rising afterglows could also be due to a structured outflow endowed with two 
"hot spots", one moving directly toward the observer and giving the prompt GRB emission, and 
another one moving slightly off the direction toward the observer \cite{granot02A}, at an angle 
$\theta_{obs} > \Gamma_0^{-1}$, its emission becoming visible when the outflow Lorentz factor 
decreases to $\Gamma = \theta_{obs}^{-1}$. The same result could be accomplished with 
an axially symmetric outflows having a bright core that yields the burst emission and a
bright ring that produced the rising afterglow when it becomes visible.
 Thus, a late-rising afterglow may be a 
relativistic effect rather than the signature of some ejecta with a lower initial Lorentz 
factor, either shock (reverse or forward) being a possible origin of the rising afterglow. 
In fact, this model is found to account better for the peak luminosity--peak time 
anti-correlation exhibited by a dozen optical afterglows with early, fast rises than the 
pre-deceleration external-shock model \cite{alin08}, although it should be noted that only
half of that correlation is real (i.e. optical peaks do not occur later and are not brighter 
than a certain linear limit in log-log space) while the other half is just an observational
bias, as there are many optical afterglows exhibiting decaying fluxes from first measurement 
that fall below the peak flux--peak time relation found for the fast-rising afterglows
(Figure \ref{rises}).   
 
\begin{figure}
    \parbox[h]{14cm}{\vspace*{-10mm} \includegraphics[height=7cm,width=14cm]{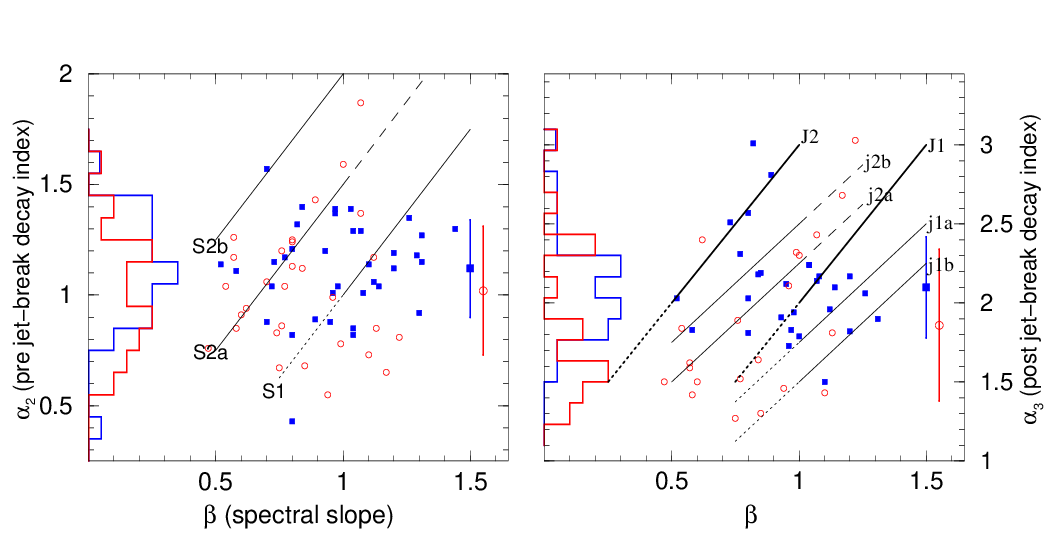} }
\caption{ \footnotesize
    Spectral slopes vs. power-law decay indices before the jet-break (left panel, 66 afterglows) 
    and after (right panel, 49 afterglows). A jet-break is defined by a light-curve steepening
    occurring after 0.1 d and to a post-decay index larger than 1.5 (with allowance for
    uncertainties, which are $\siml 0.15$ for either plotted quantity).
    Blue filled symbols are for X-ray light-curves, red open symbols are for optical light-curves,
    with same colour coding being used for the histograms at left and the average decay indices  
    shown at right (symbols with large error bars). 
    Straight lines indicate model expectations, with the following designations: 
    "1" for $\nu_c < \nu_{obs}$, "2" for $\nu_{obs} < \nu_c$, "a" for a uniform medium, 
    "b" for a wind, "j" for a conical jet (no lateral spreading), "J" for a spreading jet. 
    Dotted lines indicate a model for which the power-law index of the electron distribution 
    with energy is $p < 2$, solid for $p \in (2,3)$, dashed for $p > 3$. }
\label{ab}
\end{figure}

 Most afterglow observations were made during the decay phase, where $\alpha$ and $\beta$
are expected to be correlated and satisfy one or more closure relationships. The left 
panel of Figure \ref{ab} shows the temporal and spectral indices of optical and X-ray afterglows 
measured before the jet break and the expectations for the (post-deceleration) forward-shock 
model (as it seems more likely that the reverse shock dominates the afterglow emission only
until at most 1 ks). Surprisingly, no significant correlation can be seen between $\alpha$
and $\beta$, which may be taken either as indication that the standard forward-shock model
does not account for the diversity of afterglows (e.g. departures from its assumptions of
constant shock parameters would be required to explain the decays below the "S1" model, which 
are too slow) or that more than one variant of it realized which, combined with a small
baseline in $\alpha$ and $\beta$, requires a much larger sample to reveal the expected
underlying correlation. 

\subsection{Early optical afterglows}

 A bright optical emission arising from the reverse shock was predicted by \cite{sari99} and 
may have been observed for the first time in the optical counterpart (i.e. during the burst)
accompanying GRB 990123 \cite{akerlof99} and in the early afterglow emission following GRB 
021211, but without any further candidates until recently. Lacking a continuous, long-lived
injection of new ejecta, and because of the adiabatic cooling, the reverse-shock emission 
should be confined to the early afterglow. Then, the early afterglow emissions of GRB 990123
and 021211 being brighter than the extrapolation of the later flux and decaying faster are 
two reasons for attributing those two early optical emissions to the reverse shock. The larger
brightness (by 2.5 mag for 990123 and by 1 mag for 021211) could be explained by that the
number of ejecta electrons is, at deceleration (i.e. around burst end), larger by a factor 
$\Gamma_0 \sim 100$ than in the forward shock, and by a smaller factor at later times (as for 
021211), with some relative dimming of the reverse shock optical flux attributed to the peak 
of the reverse-shock synchrotron spectrum being lower than that of the forward shock. 

\begin{figure}
    \parbox[h]{12cm}{\includegraphics[height=5.5cm,width=12cm]{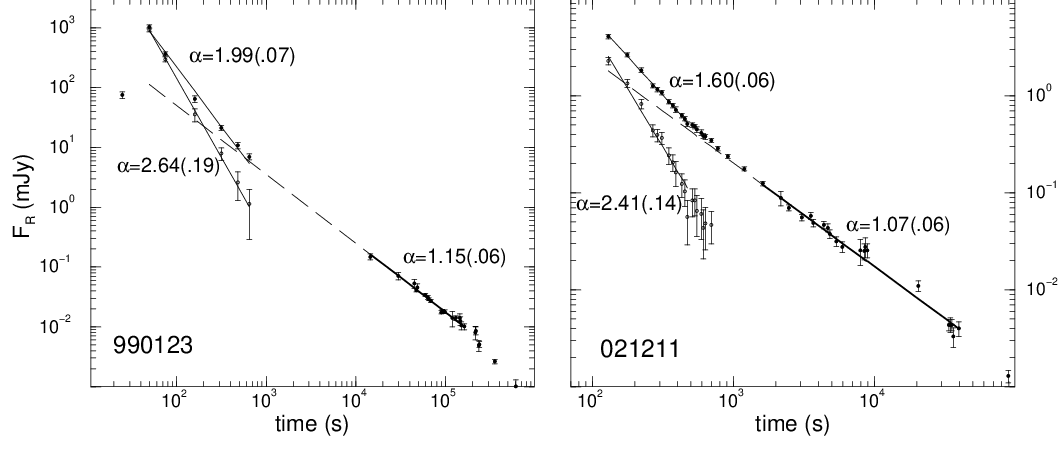} }
    \parbox[h]{4.5cm}{ \footnotesize
  Optical afterglows of GRBs 990123 and 021211, displaying a fast decay at earlier times.
  The late (after 1 ks) optical flux decay is fit with a power-law (solid, thick line), 
  extrapolated to earlier times (dashed lines), subtracted from the early emission, leading 
  to a steeper power-law decay of the "residual" optical flux before 1 ks, which could
  be attributed to the adiabatically cooling ejecta after they have been energized by the 
  reverse shock. }
\caption{ }
\label{oc}
\end{figure}

 However, both the above reasons disappear if the origin of time is not at
trigger but sometime later, e.g. at 30--40 s (corresponding to the peak of GRB 990123
optical flash). Then, the early optical emission appears as a small deviation of the 
extrapolation of the late flux and the entire afterglow light-curve is consistent with 
a single power-law, indicating a unique dissipation mechanism. Thus, absent spectral
information, the evidence for a reverse shock origin of the early optical emissions of
GRB 990123 and 021211 is circumstantial.

 Such spectral information has been acquired only recently, for the early optical
emission of GRB afterglows 061126 and 080319B. For the former, \cite{perley08} finds
that the steeper-decaying early (up to 200 s) optical emission is harder than at later 
times (after 1 ks). The indices $\alpha \simeq 2.0$ and $\beta \simeq 0.9$ of the early 
optical emission of GRB 061126 are consistent with the closure relation expected for 
adiabatic cooling ejecta. The spectral evolution observed simultaneously with the slowing 
of light-curve decay suggests the emergence of a different component after 1 ks. 

 The optical afterglow of GRB 080319B also displayed a spectral hardening simultaneous 
with the reduction in the flux decay rate \cite{wozniak08}, supporting a reverse-shock
origin of the early fast-decay phase and forward-shock origin of the later slower-decaying
emission. However, the decay at early times is too fast (for the measured spectral slope) 
to be attributed to the adiabatic cooling of ejecta. In fact, the decay index $\alpha 
\simeq 2 +\beta$ is consistent with the expectations for the large-angle emission released 
during the burst. However, that does not exclude a reverse-shock origin of the early optical 
flux, as the cooling frequency may have fallen below the optical, revealing the large-angle 
emission.

 The slow softening of the optical spectrum of GRB afterglow 080319B after 1 ks was a
surprise. If the rather flat ($\beta = 0$) spectral slope at 1 ks were due to the peak 
energy of the forward-shock synchrotron spectrum being in the optical, then a much faster 
softening is expected, given that the injection frequency evolution $\nu_i \propto t^{-3/2}$ 
is also fast. Energy injection in the blast-wave or an increasing electron/magnetic shock 
parameters could account for slow decrease of $\nu_i$ required by the slow spectral softening 
of GRB 080319B afterglow optical emission. 

\begin{figure}
    \parbox[h]{7cm}{ \includegraphics[height=5.5cm,width=7cm]{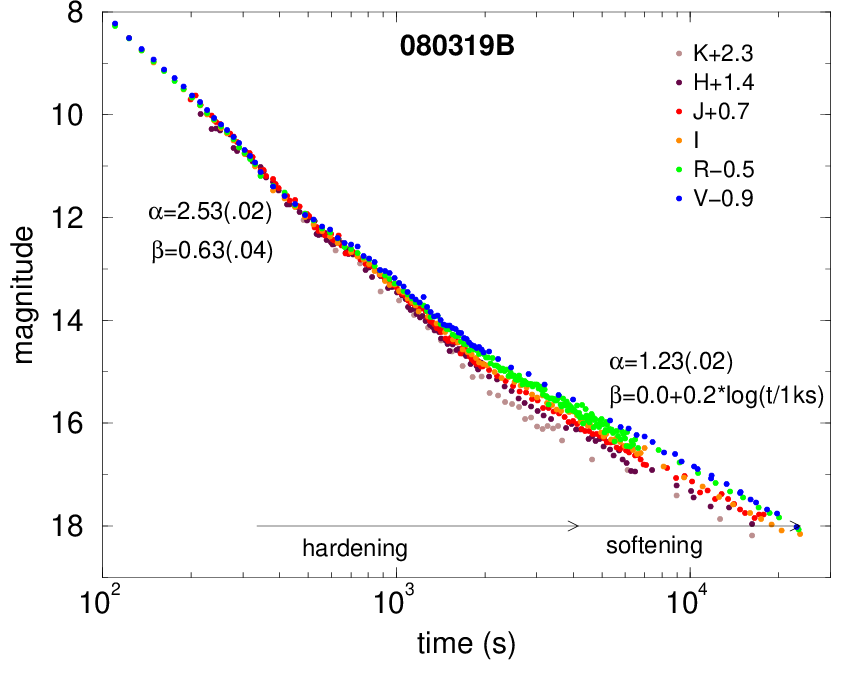} }
\hspace*{7mm}
    \parbox[h]{7cm}{ \vspace*{-5mm} \footnotesize
     Multi-colour optical observations of GRB afterglow 080319B. Light-curves have been
     shifted by indicated amounts to overlay them at earliest measurements. The fanning
     of light-curves, with redder bands appearing dimmer than bluer bands, shows a 
     hardening of the optical spectrum at 0.5--2 ks, simultaneous with the slowing of 
     the decay rate, which indicates the emergence of a different emission component.
     A slow softening of the emission from the emerging component is seen. The faster-decaying 
     early component and the slower-decaying late component are most naturally identifiable
     with the reverse and forward-shock emissions, respectively. }
\caption{}
\label{080319}
\end{figure}

\subsection{Jet-breaks}

 A tight collimation of GRB ejecta, into a jet of half-aperture less than 10 degrees, is 
desirable to reduce the isotropic-equivalent GRB output, reaching $10^{54.5}$ erg, to lower 
values, below $10^{52.5}$ erg, compatible with what the mechanisms for production of 
relativistic jets by solar-mass black-holes can yield. Besides the $\alpha-\beta$ closure
relationship being satisfied self-consistently (i.e. by a forward-shock model with same
features before and after the jet-break), {\sl achromaticity} of the break (i.e. simultaneous 
occurrence at all frequencies) is an essential test of this model. 

 The steepening of the afterglow flux decay due to collimation of ejecta was predicted by
\cite{rhoads99} and was observed for the first time in the optical emission of GRB afterglow 
990123 \cite{kulkarni99}. About 3/4 of well-monitored pre-Swift optical afterglows displayed
jet-breaks at 0.5--3 day, as shown in the compilation of \cite{zeh06}. The X-ray coverage 
of pre-Swift afterglow extended over at most 1 decade in time and was insufficient to test
for the existence of an X-ray light-curve break simultaneous with that seen in the optical.

 A smaller fraction, between 1/3 and 2/3, of Swift X-ray afterglows also exhibit jet-breaks 
\cite{alin07A}, defined as a steepening occurring after 0.1 day from a power-law decay 
with $\alpha \siml 1$ to one with $\alpha > 1.5$. Comparing the post jet-break temporal and 
spectral indices with the expectations for the forward-shock model (right panel of Figure 
\ref{ab}) shows that that model accounts for observations of post jet-break decays if jets 
are both both spreading and conical. However, just as for pre jet-break decays, the expected 
$\alpha-\beta$ correlation is not seen. 

\begin{figure}
    \parbox[h]{15cm}{ \includegraphics[height=6cm,width=15cm]{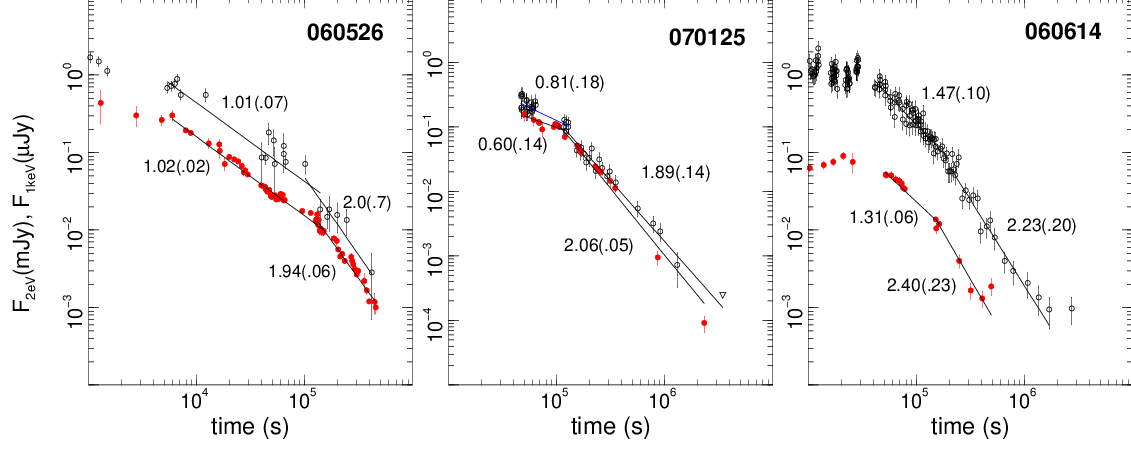} }
\caption{ \footnotesize
  Optical (red filled symbols) and X-ray (black open symbols) light-curves of three afterglows 
  displaying achromatic breaks at late times, followed by steep post-break decays, as expected 
  for a jet. Power-law decay index and uncertainty are given for each light-curve segment. }
\label{jets}
\end{figure}

 While there are many examples of potential jet-breaks in the X-ray light-curve monitored
by Swift, few are sufficiently well-monitored in the optical to test for the achromaticity
of the break. Figure \ref{jets} shows the only 3 afterglows sufficiently sampled and
followed sufficiently late to search for achromatic light-curve breaks. Besides the
simultaneity of the optical and light-curve breaks, note the equality of the pre and post-break
decay indices.

 The smaller fraction of Swift X-ray afterglows that exhibit jet-breaks, relative to that
of pre-Swift such optical afterglows, could be due to Swift detecting and localizing 
afterglows that are fainter than those followed in the optical prior to Swift. The argument
here is that, if all jets had the same energy (e.g. \cite{frail01}), then the afterglow flux 
and jet-break time should be anti-correlated: $F_\nu \propto dE/d\Omega \propto \theta_{jet}^{-2}$ 
and $t_{jet} \propto (dE/d\Omega) \theta_{jet}^4 \propto \theta_{jet}^2$ (for a wind-like medium), 
leading to $F_\nu \propto t_{jet}^{-1}$, where $dE/d\Omega$ is the forward-shock's kinetic 
energy per solid angle, thus dimmer afterglows should display later jet-breaks. 

\begin{figure}
   \parbox[h]{12cm}{ \includegraphics[height=6.5cm,width=12cm]{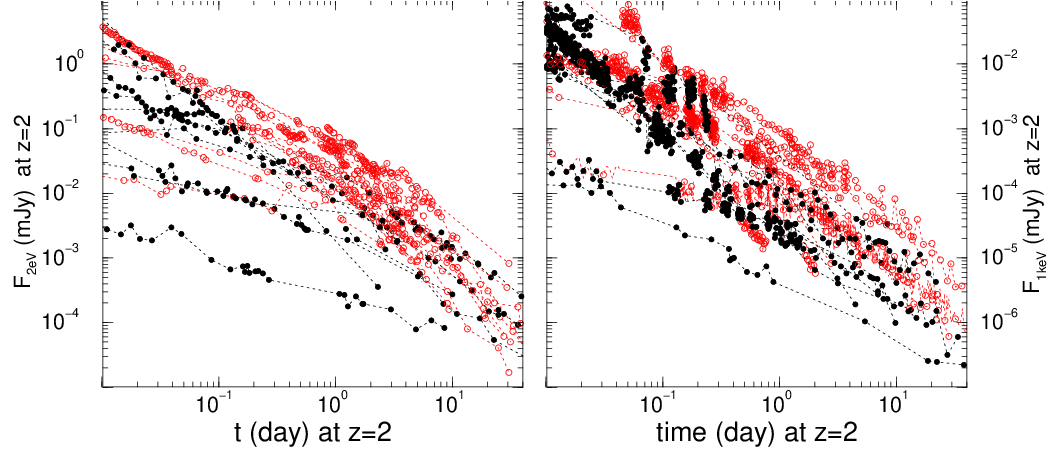} }
   \parbox[h]{4.5cm}{ \footnotesize
   Optical (left panel) and X-ray (right panel) fluxes of afterglows that displayed late 
   light-curve breaks followed by decays steeper than $\alpha=1.5$ (16 in optical, 12 in
   X-ray) or did not have a break until 10 days (12 optical, 10 X-ray), moved to same 
   redshift, to illustrate that afterglows with light-curve jet-breaks (red symbols) are 
   10 times more luminous, on average, than those without such a break (black symbols). }
\caption{ }
\label{ox}
\end{figure}

 As shown in Figure \ref{ox}, the afterglows with jet-breaks at 0.3--10 days are brighter 
by a factor $\sim 10$ than those without jet-breaks detected until 10 days, thus the
anti-correlation between afterglow flux and jet-break time expected for a universal jet
energy is confirmed, even though jet energies inferred from the timing of afterglow 
light-curve breaks have a broad distribution (e.g. \cite{ghirlanda04}). That ratio of 10
between the average brightness of afterglows with breaks and of those without breaks until 
10 days and the above-derived $F_\nu \propto t_{jet}^{-1}$ imply that the latter type of 
afterglows should display a break at 3--100 days, which could be missed if monitoring does 
not extend for sufficiently long times. For this reason, some of the dimmer X-ray afterglows 
detected by Swift may have breaks that are too late to be observed, leading to an apparent 
paucity of Swift X-ray afterglows with jet-breaks, as noted by \cite{burrows06}.

\subsection{Slowly-decaying radio afterglows}

 The long-time monitoring of radio afterglows showed that often there is an incompatibility 
between the radio and optical flux decays. 
 After the peak of the forward-shock synchrotron spectrum falls below the radio domain, 
which should happen within $\sim 10$ days and is, indeed, observed in the radio spectra 
of GRB afterglows 970508 \cite{frail00}, 021004, and 030329 (Figure \ref{radiopk}),  
the radio and optical flux decays are expected to be similar, up to a difference 
$\delta \alpha = \pm 1/4$ that could occur if the cooling break is in between radio and 
optical and if the jet is not laterally spreading.

\begin{figure}
   \parbox[h]{11cm}{ \includegraphics[height=6.5cm,width=11cm]{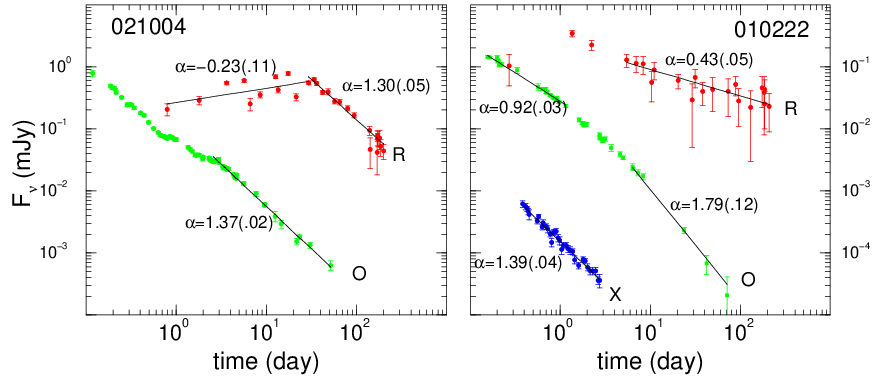} }
   \parbox[h]{5cm}{ \footnotesize
   Coupled (left panel) and decoupled (right panel) optical and radio light-curves.
   For the former, the post-peak radio decay index is the same as in the optical;
   for the latter, the radio flux decays much slower than the post jet-break optical flux.
   From top to bottom, data are radio (red symbols), optical (green) and X-ray (blue)
   measurements.  }
\caption{ }
\label{or}
\end{figure}

 In a set of nine pre-Swift afterglows with long temporal coverage at both frequencies,
I find that the above expectation is met by only four: GRB afterglows 980703, 970508, 000418,
and 021004, but that the radio flux of GRB afterglows 991208, 991216, 000301C, 000926, and 
010222 decay much slower than in the optical, with $\alpha_{opt} - \alpha_{rad} = 0.8, 0.8,
1.3, 1.6, 1.5$, respectively \cite{alin04A} (see also \cite{frail04}). For all the above 
three cases for which the optical and radio decays are well-coupled, the decays are slower 
than $\alpha = 1.5$, indicating a wide jet, while for all the five cases of decoupled radio
and optical decays, the optical displays a decay steeper than $\alpha = 1.5$ after a $\sim 1$
day break that could be interpreted as a jet-break. Thus, whenever the optical flux decays
fast, there seems to be a mechanism which produces radio emission in excess of that expected
for the forward-shock model. An example of each type of radio afterglow is shown in Figure \ref{or}.

 Because the slow radio flux decay is observed {\sl at the same time} as the faster optical 
decay, the decoupling of radio and optical light-curves cannot be attributed to energy injection
in the blast-wave, to a structured outflow, or to evolving micro-physical parameters (mechanisms
which have also been used to explain the slow early decays seen in Swift X-ray afterglow plateaus),
nor to the transition to non-relativistic dynamics. Instead, the decoupled radio and optical 
light-curve decays may indicate that these emissions arise from different parts of the 
relativistic outflow. 

 That could happen if the outflow endowed with angular structure (in the sense that 
its kinetic energy per solid angle is anisotropic), with a core that dominates the optical 
emission, yielding the jet-break, and an outer, wider envelope that produces the radio emission. 
The problem with this model \cite{alin04A} is that, optical and radio afterglows being long-lived, 
the spectral break frequencies of the core and envelope emissions evolve substantially, making 
it impossible for their emissions to be dominant over such long timescales at only one frequency, 
i.e. without "interfering" with the emission of the other part of the outflow. Shortly put, 
it is quite likely that the emission from the radio envelope would soon dominate the optical 
emission from the core and change the initially steep optical flux decay into a slower one.
(While that is a general issue for explaining decoupled afterglow light-curves with a structured
outflow, \cite{racusin08} shows that it can be avoided for GRB 080319B, whose optical and X-ray 
light-curve decays are decoupled for until 1 day). 

 Another possibility is that the optical emission arises in the forward shock while the
radio is from the reverse shock. For adiabatic cooling, the ejecta emission should decay 
slowly, even when observations are at a frequency below that of the spectral peak,
thus a reverse shock energizing the ejecta is required to account for the flat or slowly
rising part of radio light-curves (up to about 10 days). In this case, the light-curve
decay depends on the law governing the injection of fresh ejecta into the reverse shock,
the observed radio light-curve indices being close to the closure relations derived in the 
previous section for a uniform radial distribution of the incoming ejecta mass.

 In the reverse-forward shock model for afterglows with different radio and optical 
light-curve decays, the cross-interference issue may also exist, as the forward-shock
synchrotron peak flux, being larger than $\sim 0.1$ mJy at 1 day (to account for the $\sim$ 
20 magnitude optical flux), could over-shine the reverse-shock radio emission at some later
time, when the peak energy of the forward-shock emission spectrum reaches the radio domain.
This issue is best addressed with numerical calculations of the blast-wave dynamics and 
radiation. In this way, I found \cite{alin05} that the most likely solution for the 
decoupled radio and optical light-curves is that the radio afterglow emission is dominated 
by the reverse shock during the first decade in time, with the forward-shock emission peaking 
in the radio at about 100 days, overtaking that from the reverse shock sometime during the 
second decade. By itself, each component would display a decay faster than observed, but 
their sum resembles a shallow power-law over two decades in time.

\subsection{X-ray plateaus and chromatic breaks}

\begin{figure}
    \parbox[h]{7cm}{ \includegraphics[height=5.5cm,width=7cm]{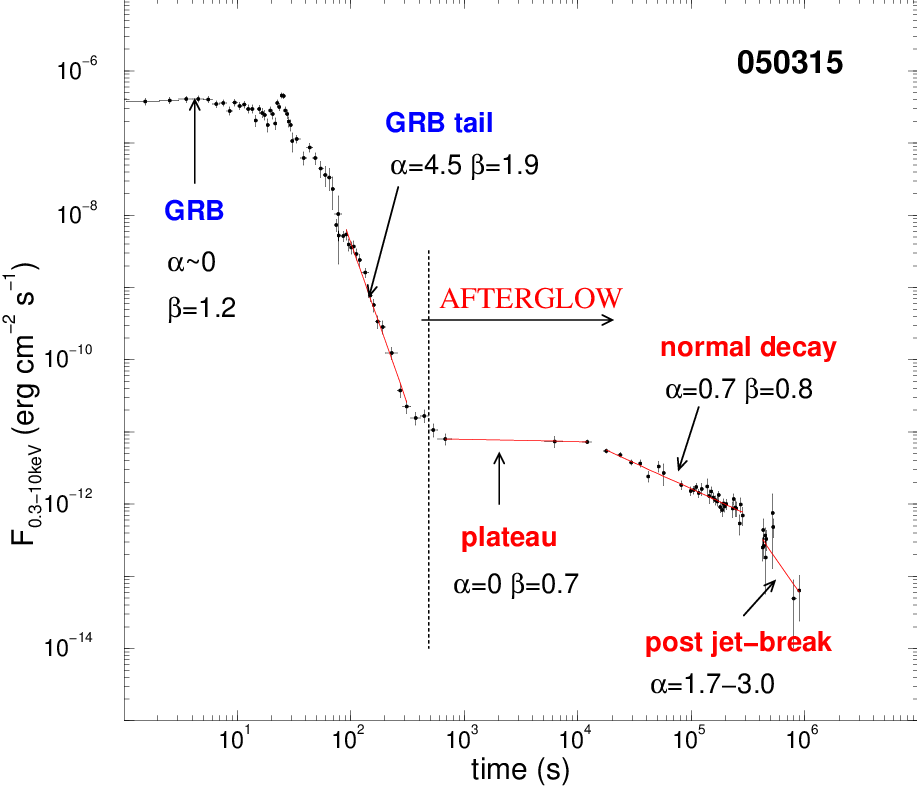} }
    \hspace*{5mm}
    \parbox[h]{5cm}{ \footnotesize
     X-ray emission during the prompt and afterglow phases of GRB 050315.
    The plateau (slow decay at 1--10 ks) is a novel feature discovered by Swift in about 
    3/4 of X-ray afterglows. }
\caption{ }
\label{0315}
\end{figure}

 The Swift satellite has opened a new temporal window for observations of X-ray afterglows,
which previously were monitored by BSAX only after 8 hours after trigger. The major surprise
(i.e. a feature not predicted) in Swift observations was that, although it appeared that the
X-ray afterglow emission at hours and days extrapolated back to the burst time would match 
the GRB flux, implying a smooth transition from counterpart to afterglow emission,
the X-ray flux from burst end to several hours is much less than that back-extrapolation,
displaying at 0.3--10 ks a phase of slow decay, with $\alpha \in (0,3/4)$. In fact, that
should have been a partial surprise because BSAX has observed a sharply decaying GRB tail 
in at least three cases, indicating that a phase of slow X-ray decay must exist at the burst 
end. Figure \ref{0315} illustrates the "plateau" phase observed for GRB afterglow 050315.

 In the simplest form of the blast-wave model, the magnitude of the light-curve decay 
steepening at the end of the plateau requires the peak of the synchrotron spectrum to 
fall below the X-ray band at the end of the plateau. However, that explanation is ruled out 
by that, observationally, the plateau end is most often not accompanied by the a spectral 
evolution \cite{nousek06,willingale07,liang07}, although exceptions exist \cite{butler07}.

 Because the plateau phase is followed by a "normal" decay, compatible with the expectations
of the standard forward-shock model, it is natural to think that departures from the assumptions
of that standard model are the cause of X-ray plateaus: 
(1) increase of the average energy per solid angle of the blast-wave area visible to the observer 
by means of 
 (1a) energy injection in the blast-wave owing to some late ejecta catching-up with the 
  forward-shock \cite{nousek06,zhang06,alin06}, or by absorbing low-frequency electromagnetic 
  radiation from a millisecond pulsar \cite{zhang06},
 (1b) an anisotropic outflow \cite{alin06,eichler06},
(2) evolving shock micro-physical parameters \cite{fan06,ioka06}, and
(3) blast-wave interacting with an "altered" ambient medium, shaped by a GRB precursor 
  \cite{ioka06}.
The effect on the afterglow flux decay of the above mechanisms for a variable "apparent" 
kinetic energy of the blast-wave were first investigated by \cite{meszaros98,rees98},
the X-ray plateaus discovered by Swift several years later providing the first tentative
confirmation that those mechanisms may be at work. 

 However, the discovery of {\sl chromatic} light-curve breaks at the end of the X-ray plateau
\cite{fan06,watson06,alin06A}, which are not seen in the optical as well, soon showed that 
neither of the above mechanisms for X-ray plateaus provide a complete picture of the afterglow 
phenomenon, as in all those models the break should be {\sl achromatic}, manifested at all
frequencies. Evolution of shock parameters for electron and magnetic field energies could
"iron out" the optical light-curve break produced by the other mechanisms listed above,
provided that the cooling frequency is between optical and X-ray (to allow a way of decoupling
the optical and X-ray light-curves), however there is no reason for their evolution to
conspire and hide the optical light-curve break so often (universality of the required 
micro-physical parameter evolutions with blast-wave Lorentz factor would provide some 
support to this contrived model).

\begin{figure}
    \parbox[h]{16cm}{ \includegraphics[height=6cm,width=16cm]{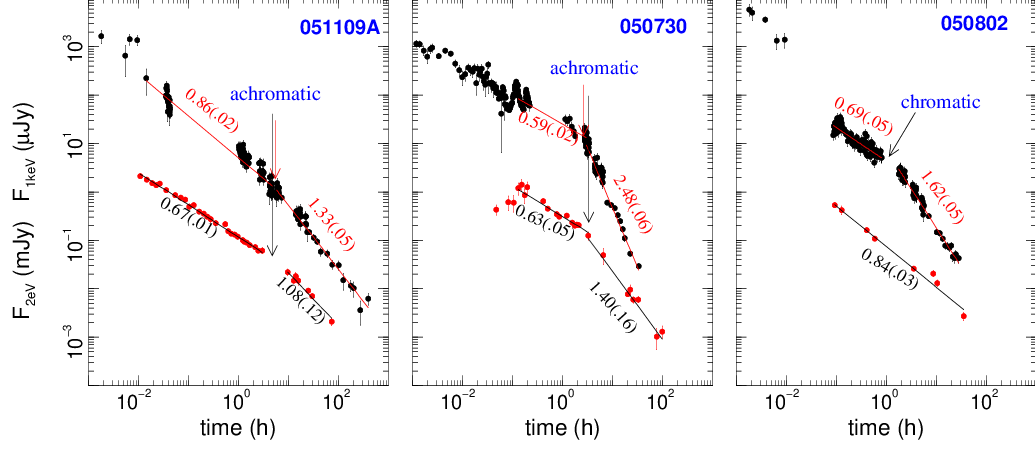} }
\caption{ \footnotesize
    Examples of an achromatic break (left panel) for which the optical and X-ray
    decay indices are similar, thus cessation of energy injection in the blast-wave
    could explain it, an achromatic break (middle panel) for which the post-break
    decay indices differ substantially, requiring that micro-physical parameters are
    also evolving, and a chromatic X-ray light-curve break (right panel) that cannot
    be attributed to an end to energy injection in the blast-wave.  }
\label{xbr}
\end{figure}

 To date, I find that there are 11 good cases of chromatic X-ray breaks, 6 good cases
of achromatic breaks, and 3 afterglows displaying well-coupled optical and X-ray light-curves:
a single power-law decay, of same decay index at both frequencies, extending over over three 
decades in time. Figure \ref{xbr} shows two examples of achromatic breaks, one with discrepant 
post-break optical and X-ray decay indices, and one chromatic X-ray break. 

 To explain the chromatic X-ray breaks, \cite{uhm07} and \cite{genet07} have proposed that 
the {\sl entire} afterglow emission is produced by the reverse shock and have shown that, by 
placing either the injection or the cooling frequency between optical and X-ray, decoupled 
light-curves can be obtained. Just as energy injection in the blast-wave, this mechanism
relies on the existence of a long-lived central engine that expels ejecta until the last
afterglow measurement (the existence of a reverse shock at days is also required by the slow 
decays seen in a couple of radio afterglows). On the other hand, the
transition observed in two optical afterglows from a fast-decaying phase to one of slower 
decay at 1 ks, accompanied by spectral evolution, argue in favour of the reverse-shock 
emission being dominant only up to 1 ks, after which it seems more natural to attribute 
the afterglow emission to the forward-shock. Furthermore, the results shown by \cite{uhm07} 
show a softening of the reverse-shock optical spectrum at the transition from fast to
slow decay, which is in contradiction with the hardening observed in GRB afterglows 061126
and 080319B. Thus, the reverse-shock model may not provide a correct description of the 
entire afterglow emission.

\begin{figure}
    \parbox[h]{7cm}{ \includegraphics[height=5.5cm,width=7cm]{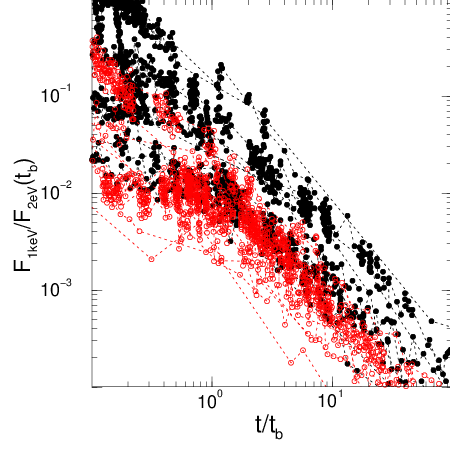} }
     \hspace*{5mm}
    \parbox[h]{9cm}{ \footnotesize \vspace*{-3mm}
     X-ray flux normalized to optical flux at the plateau end for 11 afterglows (black symbols) 
     with chromatic X-ray breaks (GRB 050401, 050607, 050802, 060605, 060927, 061121, 070110, 
     070420, 080310, 080430, 080605) and 9 afterglows (red symbols) with coupled light-curves 
     (6 achromatic breaks at plateau end: GRB 051109, 060206, 060526, 060614, 060714, 060729, and 
     3 long-lived power-law decays of same index at both frequencies: GRB 050801, 060418, 061007). 
     Time is normalized to the plateau end, occurring at $\overline{t_b} \siml 10$ ks 
     (for the 3 afterglows with single power-law decays, $t_b = 10$ ks was chosen as reference 
     epoch). Afterglows with chromatic X-ray breaks have a larger ratio $F_x/F_o(t_b)$ than 
     afterglows with coupled light-curves: $\log F_x/F_o(t_b) = -1.5 \pm 0.5$ and $-2.2 \pm 
     0.4$, respectively, indicating the existence of a mechanism which yields an X-ray 
     emission brighter by a factor $\sim 5$ than that produced by the optically-radiating 
     mechanism, for afterglows with chromatic X-ray light-curve breaks. }
\caption{ }
\label{fxfo}
\end{figure}

 On average, the X-ray to optical flux ratio is larger for afterglows with chromatic X-ray 
breaks than for afterglows with coupled optical and X-ray light-curves (i.e. with achromatic 
breaks or single power-law decays), as shown in Figure \ref{fxfo}. This indicates that 
chromatic X-ray breaks are due to a mechanism whose emission over-shines an underlying 
one only in the X-rays, but not in the optical. Thus, it seems that the diversity of 
optical vs. X-ray light-curve behaviours should be attributed to the existence of two 
mechanisms for afterglow emission and not to a unique origin. 

 So far, three proposals along that line have been put forth: dust-scattering of the blast-wave 
emission, bulk and inverse-Compton scattering of the same emission, and a central-engine 
mechanism that could be the same internal shocks in a variable wind that are believed to 
produce the prompt burst emission \cite{rees94}. 
 
 \cite{shao07} have proposed that X-ray plateaus result from scattering by dust in 
the host galaxy, much like the expanding rings produced by dust-scattering in our 
Galaxy in GRB afterglows 031203 \cite{vaughan04} and 050713A \cite{tiengo06}. However, 
for dust-scattering, harder photons are those scattered at a smaller angle, thus they 
arrive earlier at observer, leading to a strong spectral softening of the X-ray light-curve 
plateau, of $\delta \beta \simeq 2$, and to a strong dependence of the plateau duration 
on the photon energy, $\Delta t \propto \nu^{-2}$, both of which are clearly refuted 
by afterglow observations \cite{shen08}.

 \cite{ghisellini08} have proposed that, in some afterglows, the "central engine" makes
a substantial contribution to the afterglow X-ray flux. This model requires a central engine 
that operates until the last afterglow detection; the dissipation mechanism may be shocks 
in a variable outflow, which can also account for the bright and short-lived flares observed 
in many Swift X-ray afterglows (e.g. \cite{burrows07,chincarini07}). 

 Given that the forward-shock model with cessation of energy injection at the plateau end 
can explain the achromatic breaks and that the standard forward-shock model accounts for 
the coupled single power-law light-curves, it would be more desirable to identify a mechanism 
for producing decoupled X-ray and optical light-curves that is related to the forward shock 
and which dominates its emission only occasionally. 

 Bulk and inverse-Compton scattering of the forward-shock photons by an outflow interior 
to the blast-wave is such a mechanism. In this model \cite{alin08A}, all the afterglow 
emission originates
in the forward shock, which explains so naturally the long-lived, power-law decay of GRB
afterglows, coupled optical and X-ray light-curves resulting when the scattered emission
is dimmer than the forward-shock's, while chromatic X-ray light-curve breaks occur when 
the scattered emission is dominant in the X-ray. For this model to work, the scattering
outflow must be almost purely leptonic, to ensure a sufficiently high (sub-unity) optical 
depth to electron scattering to account for the observed X-ray flux, assuming that the
kinetic energy of the scattering outflow is not much larger than that of the forward shock.
The same outflow also injects energy into the blast-wave and, if that energy is larger
than the forward shock's, then it mitigates the blast-wave deceleration, producing a 
light-curve plateau ending with an achromatic break when the injected energy falls below
that of the forward shock and stops being dynamically important. Therefore, a delayed
outflow is the origin of both chromatic and achromatic light-curve breaks, the former
occurring when the scattered emission is dominant, while the latter happening when the
forward-shock emission is dominant. Still, the achromatic break of GRB 050730 (Figure
\ref{xbr}), followed by an X-ray flux decay much steeper than that of the optical cannot
be explained with only cessation of energy injection at the time of the break, and requires
an extra feature, that the shock micro-physical parameters are not constant. 

 The above scattering model also explains late X-ray flares, which arise from dense or
hot (i.e. with relativistic electrons) sheets within the outflow. When dominant, the
scattered X-ray emission received at time $t$ reflects the properties (density, Lorentz 
factor) of the outflow at $ct/(z+1)$ behind the forward shock, sharp drops of the X-ray
flux as that observed for GRB afterglow 070110 at 30 ks being due to a gap in the 
scattering outflow.

  For an instantaneous release 
of all the ejecta, the kinematics of outflow radial-spreading owing to different initial 
Lorentz factors, followed by deceleration of the forward-shock, leads to that, when the 
ejecta of Lorentz factor $\Gamma_{sc}$ catch up with the forward shock, moving at $\Gamma_{fs}$, 
the Lorentz factor contrast is $\Gamma_{sc}/\Gamma_{fs} = \sqrt{4-s}$ (i.e. 2 for a 
homogeneous medium and $\sqrt{2}$ for a wind). That ratio is too small for bulk-scattering
to boost enough the forward-shock emission to dominate that arriving directly from the
forward shock. But, if the scattering outflow was energized by internal shocks, 
inverse-Compton scatterings by relativistic electrons (of comoving frame energy $\gamma_e
m_e c^2$) can achieve that goal. In fact the properties of the scattered emission depend 
only on the product $\Gamma_{sc} \gamma_e$. Thus, for a sudden release of ejecta to lead 
to a sufficiently bright scattered emission, the scattering outflow should be hot. 
Alternately, if the scattering outflow is cold, then the larger ratio $\Gamma_{sc}/\Gamma_{fs}
\simg 100$ necessary for the scattered flux to over-shine that from the forward shock
requires a long-lived engine.

 As a general test of all models that explain decoupled optical and X-ray afterglows by
attributing them to different mechanisms, there should not be any afterglows whose
optical and X-ray light-curves evolve from decoupled (i.e. with a chromatic break) to 
coupled (i.e. with an achromatic break), or vice-versa, unless one of the light-curves 
displays a sudden flux or spectral change that would indicate a second mechanism becoming 
dominant. So far, I find only two cases of afterglows whose optical and X-ray light-curves 
evolve from decoupled to coupled (GRB afterglows 070110 and 080319B), but their X-ray 
light-curves display, indeed, a sharp drop (at 20 ks, in both cases).

\section{Conclusions}

 The temporal and spectral properties of GRB afterglows are, in general, consistent with
those predicted for the synchrotron emission from the blast-wave produced when highly 
relativistic ejecta (initial Lorentz factor above 100) interact with the ambient medium.

 As expected from shocks accelerating particles with a power-law distribution with energy, 
power-laws are observed in the optical and X-ray afterglow continua.
A spectral softening (i.e. decrease of peak frequency) is expected owing to the blast-wave 
deceleration and is observed in radio afterglows spectra and in the behaviour of afterglow 
radio light-curves, which rise slowly until the synchrotron spectrum peak reaches the radio 
domain and fall-off afterward.

 Rising afterglow light-curves are seen at early times in the optical, and are consistent 
with the pre-deceleration emission from the forward shock, although a structured outflow 
with a hot-spot that gradually becomes visible to the observer is also possible, in which 
case the rising afterglow emission could also be explained with the reverse shock.

 Much more often, afterglow light-curves display power-law decays of indices that are
not correlated with the spectral slopes, as would be expected for the external-shock model. 
That inconsistency could be due to a substantial intrinsic scatter in decay indices and 
spectral slopes, owing to more than one variant of the blast-wave model occurring in GRB 
afterglows, combined with a small range of those indices being realized.
 
 Bright optical flashes accompanying the burst emission were predicted to arise from
the reverse shock, owing to the larger number of ejecta electrons than in the forward
shock. Fast-falling optical light-curves have been observed at 100--1000s
in two afterglows (991023 and 021211), followed by a slower of the decay, which was taken 
as evidence for the reverse-shock emission dominating the early afterglow, although
spectral information was not available to test that hypothesis. More recently, the 
early optical spectral slopes were measured for two afterglows (061126 and 080319B),
the decay of the former being consistent with that from adiabatically-cooling ejecta,
while the later is faster than expected and consistent with it being the large-angle
emission released at an earlier time.

 For the above two optical afterglows with spectral information at early times, the slowing 
of the optical flux decay at 1 ks is accompanied by a spectral evolution, which indicates 
the transition from one mechanism to another. Most naturally, that is the transition from 
ejecta emission to forward-shock emission, with the reverse-shock emission being relevant
for the optical afterglow only during its early phase.

 Evidence for a reverse-shock emission is also provided by the slow radio flux decays
observed after 10 day in several afterglows. Adiabatically cooling ejecta would yield a
decay faster than observed, particularly if the synchrotron cooling frequency were to fall
below the radio, thus a reverse-shock accelerating ejecta electrons is required to operate
for days and produce a radio emission decaying much slower than the optical at the
same time, the latter being attributed to the forward shock. Together with the above
conclusion regarding the contribution of the reverse shock to the early optical afterglow,
this suggests that the reverse shock is the main afterglow source for a duration that 
decreases with observing frequency, perhaps never being dominant in the X-rays and having
no connection with the chromatic X-ray light-curve breaks seen at $\sim 10$ ks in most
afterglows. 

 That the X-ray-to-optical flux-ratio is larger (by a factor 5) for afterglows with 
chromatic X-ray light-curve breaks than for those with coupled optical and X-ray 
light-curves (i.e. with achromatic breaks or similar power-law decays), indicates the
existence of a different mechanism producing the X-ray emission of afterglows with chromatic 
X-ray breaks, coupled light-curves resulting when the emission from that novel mechanism
is negligible. Long-lived internal shocks or scattering of the blast-wave emission by an
outflow located behind it are two possibilities that could explain chromatic X-ray breaks, 
as well as the flares seen in many X-ray light-curves. The latter mechanism is also related 
to energy injection in the blast-wave, whose cessation accounts naturally for achromatic 
light-curve breaks. Thus, in the scattering model, the diversity of optical and X-ray 
light-curve relative behaviours is attributed to the interplay between the scattered 
and direct blast-wave emissions, combined with the changing dynamics of the blast-wave
produced when the scattering outflow brings into the shock more energy than already existing.

 On energetic grounds, GRB outflows should be collimated into jets narrower than
10 degrees. The steepening of the afterglow flux decay when the jet boundary
becomes visible to the observer was another major prediction confirmed by observations.
Just as for the pre jet-break phase, more than one jet model is required to account for the
measured decay indices (given the observed spectral slopes). Swift X-ray afterglows display
light-curve jet-breaks less often than pre-Swift optical afterglows, which could be due 
to that the former afterglows (being dimmer) arise from wider jets whose jet-breaks
occur later and could, thus, be missed more often.

\end{document}